\documentclass[preprint,nofootinbib]{revtex4}%
\usepackage{amssymb}
\usepackage{amsfonts}
\usepackage{amsmath}
\usepackage{graphicx}%
\providecommand{\U}[1]{\protect\rule{.1in}{.1in}}
\begin{document}
\title{Asymptotically Lifshitz wormholes and black holes\\for Lovelock gravity in vacuum}
\author{Javier Matulich$^{1,2}$ and Ricardo Troncoso$^{1}$}
\affiliation{$^{1}$Centro de Estudios Cient\'{\i}ficos (CECs), Casilla 1469, Valdivia, Chile,}
\affiliation{$^{2}$Departamento de F\'{\i}sica, Universidad de Concepci\'{o}n, Casilla,
160-C, Concepci\'{o}n, Chile.}
\preprint{CECS-PHY-11/03}

\begin{abstract}
Static asymptotically Lifshitz wormholes and black holes in vacuum are shown
to exist for a class of Lovelock theories in $d=2n+1>7$ dimensions, selected
by requiring that all but one of their $n$ maximally symmetric vacua are AdS
of radius $l$ and degenerate. The wormhole geometry is regular everywhere and
connects two Lifshitz spacetimes with a nontrivial geometry at the boundary.
The dynamical exponent $z$ is determined by the quotient of the curvature
radii of the maximally symmetric vacua according to $n(z^{2}-1)+1=\frac{l^{2}%
}{\mathcal{L}^{2}}$, where $\mathcal{L}^{2}$ corresponds to the curvature
radius of the nondegenerate vacuum. Light signals are able to connect both
asymptotic regions in finite time, and the gravitational field pulls towards a
fixed surface located at some arbitrary proper distance to the neck. The
asymptotically Lifshitz black hole possesses the same dynamical exponent and a
fixed Hawking temperature given by $T=\frac{z}{2^{z}\pi l}$. Further analytic
solutions, including pure Lifshitz spacetimes with a nontrivial geometry at
the spacelike boundary, and wormholes that interpolate between asymptotically
Lifshitz spacetimes with different dynamical exponents are also found.

\end{abstract}
\maketitle

\section{Introduction}

Exotic gravitational configurations, when consistent, naturally attract the
attention of theoretical physicists. Wormhole solutions, describing handles in
the spacetime topology, barely fall within this category. Indeed, as pointed
out by Morris, Thorne and Yurtsever \cite{Morris:1988tu, Morris:1988cz},
static wormholes in General Relativity necessarily lead to the violation of
the null energy condition localized around the neck (for a nice review see
Ref. \cite{Visser}), and the picture does not change neither in higher
dimensions nor in the presence of cosmological constant (see e.g.
\cite{Lemos:2003jb, Cataldo:2011zn}). However, in higher dimensions, the
presence of terms with higher powers in the curvature provided by certain
class of Lovelock theories, allows to circumvent this obstacle even in vacuum
\cite{DOT5,DOT6,DOT-Higher}. Thus, the possibility of violating energy
conditions is then clearly removed since the whole spacetime is devoid of any
kind of stress-energy tensor. Further wormholes solutions in Lovelock theories
with matter fields that do not conflict with energy conditions have been found
in Refs. \cite{Bhawal:1992sz, Maeda:2008nz, Thibeault:2005ha, Richarte:2007zz,
Dehghani:2009zza, Canfora:2008ka}. Besides, spacetimes with unusual asymptotic
behaviour, possessing anisotropic scaling symmetries at infinity, can be
obtained from General Relativity once endowed with \textquotedblleft
unfamiliar and contrived\textquotedblright\ matter fields (see e.g.
\cite{Balasubramanian:2009rx}). Although they obstruct the possibility of
defining all the possible conserved charges and stress-energy fluxes as in the
case asymptotically maximally symmetric spacetimes, they become relevant due
to their potential applications aimed to describe condensed matter models in
the strong coupling regime along the lines of the AdS/CFT correspondence (see
e.g. \cite{Herzog, Hartnoll, McGreevy, Maldacena gauge-gravity}). A concrete
example in this vein was first provided by the so-called Lifshitz spacetimes
in Ref. \cite{KLM} (see also \cite{Koroteev-Libanov}), and thereafter a wide
class of asymptotically Lifshitz solutions have been found, either analytic
\cite{Balasubramanian:2009rx,ALBHs0,Pang,ALBHs1,ALBHs2,Eloy,ALBHs3,ALBHs4,ABGGH,Dehghani-Mann,DML,ALBHs6,ALBHs7,ALBHs8,ALBHs9,GTT,MS}
or numerical \cite{NLBHs1,NLBHs2,NLBHs3,NLBHs4,NLBHs6,NLBHs7,Mann-numerical}.

\bigskip

One of the main results reported here is that certain class of Lovelock
theories admits exact analytic solutions exhibiting at once all of the unusual
features described above, i.e., asymptotically Lifshitz wormholes in vacuum.

\bigskip

Hereafter we will focus on Lovelock theories in $d=2n+1\geq5$ dimensions,
selected by requiring that all but one of their $n$ maximally symmetric vacua
are AdS spacetimes of radius $l$ and degenerate. In the absence of torsion,
the field equations can then be written as (see e.g. \cite{HDG})%
\begin{equation}
\mathcal{E}_{a}:=\epsilon_{aa_{2}a_{3}\cdot\cdot\cdot a_{d}}\bar{R}%
^{a_{2}a_{3}}\cdot\cdot\cdot\bar{R}^{a_{d-3}a_{d-2}}\left(  R^{a_{d-1}a_{d}%
}+\frac{1}{\mathcal{L}^{2}}e^{a_{d-1}}e^{a_{d}}\right)  =0\ ,
\label{Field Eqns}%
\end{equation}
where $\bar{R}^{ab}:=R^{ab}+\frac{1}{l^{2}}e^{a}e^{b}$, and $\mathcal{L}^{2}$
stands for the radius of the non degenerate vacuum whose sign is not fixed a
priori. Here $R^{ab}=d\omega^{ab}+\omega_{\text{\ }c}^{a}\omega^{cb}$ is the
curvature $2$-form for the spin connection $\omega^{ab}$, and $e^{a}=e_{\mu
}^{a}dx^{\mu}$ is the vielbein.

The plan of the paper is as follows. In the next section the asymptotically
Lifshitz wormhole solution is discussed, including some of their causal and
geometrical properties. An asymptotically Lifshitz black hole with the same
dynamical exponent is found in section \ref{Black hole}. Further analytic
solutions, including pure Lifshitz spacetimes with a nontrivial geometry at
the spacelike boundary, and wormholes that interpolate between asymptotically
Lifshitz spacetimes with different dynamical exponents are described in
section \ref{Further solutions}. Section \ref{Summary and discussion} is
devoted to the summary and discussion of the results, and an appendix that
concerns with a subset of the theories defined by Eq. (\ref{Field Eqns}) is
also included.

\section{Asymptotically Lifshitz wormholes in vacuum}

\label{Wormholes}

The field equations (\ref{Field Eqns}) admit the following exact solution%
\begin{equation}
ds^{2}=l^{2}\left[  -\cosh^{2}(z(\rho-\rho_{0}))dt^{2}+d\rho^{2}+\cosh
^{2}(\rho)d\Sigma_{d-2}^{2}\right]  \ , \label{Lifshitz wormhole}%
\end{equation}
where $z$ is determined by the quotient of the curvature radii of the
maximally symmetric vacua according to
\begin{equation}
n(z^{2}-1)+1=\frac{l^{2}}{\mathcal{L}^{2}}\ , \label{dynamical exponent}%
\end{equation}
and $\rho_{0}$ is an integration constant\footnote{Making $z\rightarrow-z$
just amounts to a total reflection in the radial coordinate.}. The coordinates
range as $-\infty<t<\infty$, $-\infty<\rho<\infty$, and the line element
$d\Sigma_{d-2}^{2}$, being independent of $t$ and $\rho$, stands for the
metric of the base manifold $\Sigma_{d-2}$ which is assumed to be compact and
smooth. The spacetime (\ref{Lifshitz wormhole}) is geodesically complete and
regular everywhere. It also possesses two disconnected boundaries so that it
describes a static wormhole with a neck of radius $l$ located at $\rho=0$. As
one approaches to both asymptotic regions, i.e., at $\rho\rightarrow\pm\infty
$, the metric (\ref{Lifshitz wormhole}) reads%
\begin{equation}
ds^{2}\rightarrow\frac{l^{2}}{4}\left[  -e^{2z\rho}dt^{2}+4d\rho^{2}+e^{2\rho
}d\Sigma_{d-2}^{2}\right]  \ , \label{Lifshitz rho}%
\end{equation}
which under the coordinate transformation defined through $r=\frac{l}%
{2}e^{\rho}$, $\tau=lt$, explicitly acquires the form of a Lifshitz spacetime
in Schwarzschild-like coordinates, given by\footnote{Spacetimes whose
asymptotic region approaches (\ref{Pure Lifshitz}), such that the metric of
the spacelike boundary is not flat, were recently dubbed as \textquotedblleft
asymptotically locally Lifshitz\textquotedblright\ in \cite{Ross:2011gu}.}%

\begin{equation}
ds^{2}\rightarrow-\frac{r^{2z}}{l^{2z}}d\tau^{2}+\frac{l^{2}}{r^{2}}%
dr^{2}+r^{2}d\Sigma_{d-2}^{2}\ ,\label{Pure Lifshitz}%
\end{equation}
possessing anisotropic scaling symmetry of the form $\tau\rightarrow
\lambda^{z}\tau$, $r\rightarrow\lambda^{-1}r$, provided the metric of the
spacelike boundary conformally rescales as $d\Sigma_{d-2}^{2}\rightarrow
\lambda^{2}d\Sigma_{d-2}^{2}$.

\begin{figure}[h!]
\centering
\includegraphics[angle=0,width=18 cm]{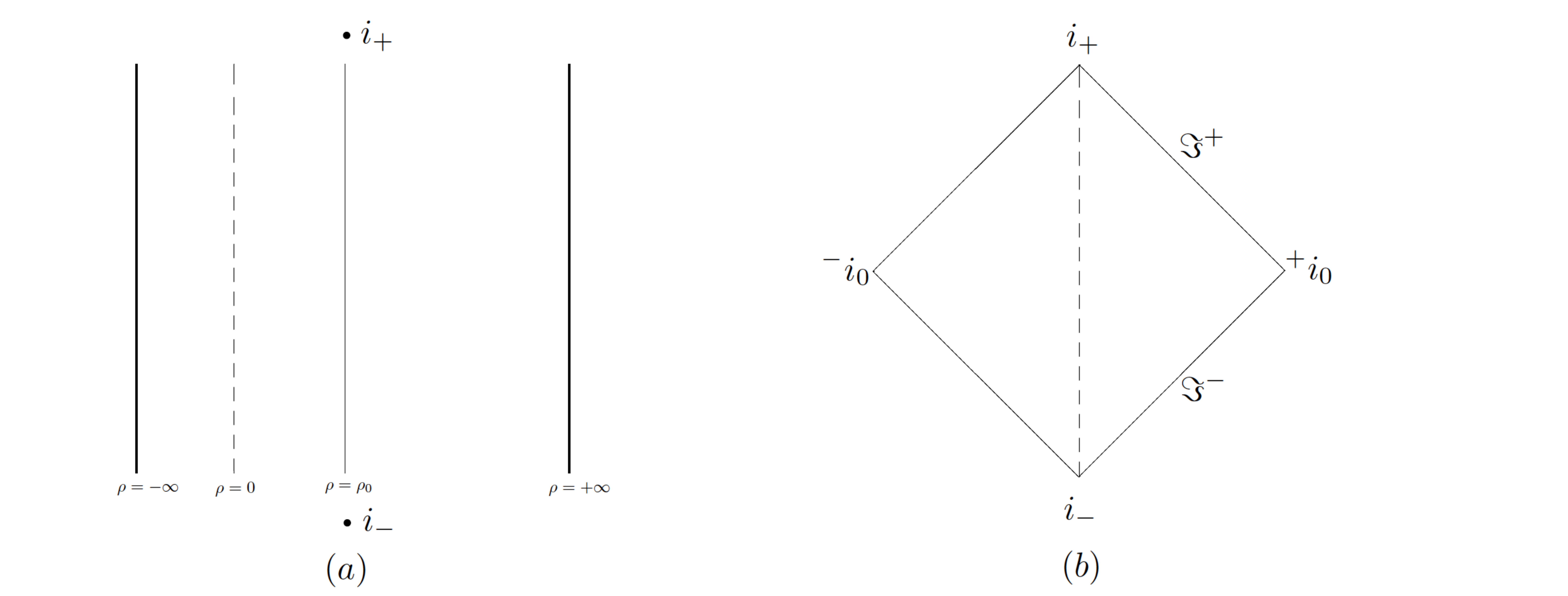}
\caption{Causal structure for the wormhole. Figs. (a) and (b) correspond to the cases $z>0$ and $z=0$, respectively.}%
\label{Fig}
\end{figure}

Therefore, the wormhole connects two Lifshitz spacetimes of dynamical exponent
$z$ with a nontrivial geometry at the spacelike boundary. For $z>0$ the causal
structure is similar to the one of AdS spacetime in two dimensions (see Fig.
\ref{Fig}a).

Note that when all the maximally symmetric vacua coincide, i.e., for
$\mathcal{L}^{2}=l^{2}$, according to Eq. (\ref{dynamical exponent}) the
dynamical exponent is given by $z=1$ and one then recovers the asymptotically
AdS wormhole solution found in Ref. \cite{DOT-Higher}. In this sense, the
dynamical exponent $z$ measures the deviation of the non degenerate maximally
symmetric vacuum with respect to the $(n-1)$-degenerate AdS ones.

One can also see that the wormhole metric (\ref{Lifshitz wormhole}) shares
many properties with its asymptotically AdS cousin. For instance, it is simple
to check that timelike curves can go forth and back from the neck, and it is
amusing to verify that radial null geodesics are able to connect both
asymptotic regions in finite time. Indeed, as it can be seen from
(\ref{Lifshitz wormhole}), a photon that travels radially from one asymptotic
region to the other, i.e., starting from $\rho=-\infty$ towards $\rho=+\infty
$, performs the entire trip in a coordinate time given by%
\[
\Delta t=\int_{-\infty}^{+\infty}\frac{d\rho}{\cosh\left(  z(\rho-\rho
_{0})\right)  }=\frac{2}{z}\left[  \arctan\left(  e^{\rho-\rho_{0}}\right)
\right]  _{-\infty}^{+\infty}=\frac{\pi}{z}\ ,
\]
which does not depend on $\rho_{0}$. A static observer located at $\rho
=\rho_{0}$, who actually stands on a timelike geodesic, then says that this
occurred in a proper time given by $\pi lz^{-1}$. Furthermore, perturbations
along $\rho$ makes this observer to oscillate around $\rho=\rho_{0}$, which
means that gravity is pulling towards the fixed surface defined by $\rho
=\rho_{0}$, being at finite proper radial distance of the neck. Indeed, radial
timelike geodesics are confined since they fulfill%
\begin{align}
\dot{t}-\frac{E}{l^{2}\cosh^{2}(z(\rho-\rho_{0}))}  &  =0\text{\ },\\
l^{2}\dot{\rho}^{2}-\frac{E^{2}}{l^{2}\cosh^{2}(z(\rho-\rho_{0}))}+\sigma &
=0\ ,
\end{align}
where dot stands for derivatives with respect to proper time, the velocity is
normalized as $u_{\mu}u^{\mu}=-\sigma$, and the integration constant $E$
corresponds to the energy. Note that the position $\rho(\tau)$ of a radial
geodesic, in proper time behaves as a particle in a P\"{o}schl-Teller
potential. It also follows that null and spacelike radial geodesics connect
both asymptotic regions in finite time.

The solution (\ref{Lifshitz wormhole}) is well defined provided%
\begin{equation}
z^{2}\geq0\ \rightarrow\ \frac{l^{2}}{\mathcal{L}^{2}}\geq1-n\ ,
\label{existence z}%
\end{equation}
and according to the range of $z$ different remarks are worth to be pointed out:

\bigskip

$\bullet$ $z^{2}>1:$ This case becomes relevant within the context of
non-relativistic holography (see e.g. \cite{ALBHs0}), since the effective
speed of light of the --interacting-- dual theories at each boundary goes to
infinity. For the wormhole in eq. (\ref{Lifshitz wormhole}), this condition is
fulfilled provided $0<\mathcal{L}^{2}<l^{2}$, which means that a
non-relativistic dual picture could be obtained if the theory described by
(\ref{Field Eqns}) admits a single nondegenerate vacuum being AdS spacetime of
radius $\mathcal{L}$ necessarily smaller than the one of the degenerate AdS
vacua, given by $l$.

\bigskip

$\bullet$ $z^{2}=1:$ As mentioned above, in this case the wormhole is
asymptotically AdS and reduces to the solution found in Ref. \cite{DOT-Higher}%
. Since $\mathcal{L}^{2}=l^{2}$, the theory admits a unique maximally
symmetric AdS vacuum \cite{BH-Scan}, and the Lagrangian can be written as a
Chern-Simons theory for the AdS\ group \cite{Chamseddine:1989nu}. Furthermore,
one can verify that the wormhole also provides a solution for the
corresponding locally supersymmetric extensions in five
\cite{Chamseddine:1990gk} and higher odd dimensions
\cite{Troncoso:1997va,Troncoso:1998ng}.

\bigskip

$\bullet$ $1-\frac{1}{n}<z^{2}<1:$ For this range the theory has to admit a
single nondegenerate vacuum being AdS spacetime of radius $\mathcal{L}$
necessarily of greater radius than the one of the degenerate AdS vacua of
radius $l$, i.e. $\mathcal{L}^{2}>l^{2}$.

\bigskip

$\bullet$ $z^{2}=1-\frac{1}{n}:$ In this case the nondegenerate maximally
symmetric vacuum of the theory has vanishing curvature and it is so given by
Minkowski spacetime ($\mathcal{L}^{2}\rightarrow\infty$). Therefore, the
volume -- or cosmological-- term in the action, proportional to $\sqrt{-g}$,
has to be absent in the action.

\bigskip

$\bullet$ $0<z^{2}<1-\frac{1}{n}:$ This range requires the theory described by
(\ref{Field Eqns}) to admit a single nondegenerate dS vacuum, fulfilling
$\mathcal{L}^{2}<\frac{l^{2}}{1-n}<0$. It is worth mentioning that for the
special values of the dynamical exponent, given by%

\begin{equation}
z^{2}=1-\frac{1}{n-k}\ , \label{Special zs}%
\end{equation}
the $k$-th power of the curvature is absent in the field equations, and hence
in the action (see Appendix). Note that the previous case, $z^{2}=1-\frac
{1}{n}$, is then consistently recovered for $k=0$.

\bigskip

$\bullet$ $z^{2}=0:$ In this case the theory (\ref{Field Eqns}) also admits a
single nondegenerate dS vacuum, whose curvature radius is fixed as
$\mathcal{L}^{2}=\frac{l^{2}}{1-n}$ and, according to (\ref{Special zs}), is
such that the $k=(n-1)$-th power of the curvature is absent in the field
equations. The metric (\ref{Lifshitz wormhole}) acquires a very simple form,
that reads%
\begin{equation}
ds^{2}=l^{2}\left[  -dt^{2}+d\rho^{2}+\cosh^{2}(\rho)d\Sigma_{d-2}^{2}\right]
\ ,\label{wormhole z=0}%
\end{equation}
connecting two static universes at $\rho\rightarrow\pm\infty$. Its causal
structure looks like the one of Minkowski spacetime in two dimensions (see
Fig. \ref{Fig}b) \footnote{Wormholes in vacuum, similar in form as compared
with eq. (\ref{wormhole z=0}), have been previously found for conformal
gravity in \cite{Oliva:2009hz} and for different three-dimensional gravity
theories in Refs. \cite{Oliva:2009zz} and \cite{OTT}. This is also the case
for the Einstein-Gauss-Bonnet theory with matter \cite{Maeda:2008nz}, as well
as for compactified Lovelock theories in eight dimensions
\cite{Canfora:2008ka}.}.

\subsection{Nontrivial spacelike boundary geometries}

\label{Base maniflod geometry}

Finding the explicit form of the base manifold metric $\Sigma_{d-2}$, which
determines the geometry of the neck as well as the one of the spacelike
boundary is not a simple task. The conditions that the metric $d\Sigma
_{d-2}^{2}$ has to fulfill can be obtained as follows:

For the metric given by eq. (\ref{Lifshitz wormhole}) the vielbein can be
chosen as%
\[
e^{0}=l\ \cosh(z(\rho-\rho_{0}))dt\ ;\ e^{1}=l\ d\rho\ ;\ e^{m}=l\cosh
(\rho)\ \tilde{e}^{m}\ ,
\]
where $\tilde{e}^{m}$ is the vielbein of $\Sigma_{d-2}$. The nonvanishing
components of $\bar{R}^{ab}=R^{ab}+\frac{1}{l^{2}}e^{a}e^{b}$ then read%
\begin{align}
\bar{R}^{01} &  =\frac{1-z^{2}}{l^{2}}e^{0}\wedge e^{1}\ ,\nonumber\\
\bar{R}^{0m} &  =\frac{1}{l}\left[  \cosh(\rho)-z\sinh(\rho)\tanh(z(\rho
-\rho_{0}))\right]  e^{0}\wedge\tilde{e}^{m}\ ,\nonumber\\
\bar{R}^{mn} &  =\tilde{R}^{mn}+\tilde{e}^{m}\wedge\tilde{e}^{n}\ ,\nonumber
\end{align}
where $\tilde{R}^{mn}$ stands for the curvature two-form of $\Sigma_{d-2}$.
Then, the component $\mathcal{E}_{0}=0$ of the field equations
(\ref{Field Eqns}) reduces to the following scalar condition on $\Sigma_{d-2}%
$:%
\begin{equation}
(z^{2}-1)\epsilon_{m_{3}m_{4}\cdot\cdot\cdot m_{d}}\bar{R}^{m_{3}m_{4}}%
\cdot\cdot\cdot\bar{R}^{m_{d-2}m_{d-1}}\tilde{e}^{m_{d}}=0\ .\label{E0}%
\end{equation}
Analogously, the combination $\mathcal{E}_{0}e^{0}-\mathcal{E}_{1}e^{1}=0$
gives a different scalar condition, which reads%
\begin{align}
\epsilon_{m_{3}\,\cdot\cdot\cdot\,m_{d}}  & \left[  n\bar{R}^{m_{3}m_{4}}%
\cdot\cdot\cdot\bar{R}^{m_{d-2}m_{d-1}}+\right.  \nonumber\\
& \left.  (n-1)(z^{2}-1)\cosh^{2}(\rho)\bar{R}^{m_{3}m_{4}}\cdot\cdot\cdot
\bar{R}^{m_{d-4}m_{d-3}}\tilde{e}^{m_{d-2}}\tilde{e}^{n_{d-1}}\right]
\tilde{e}^{m_{d}}=0\ ,\label{Eo-E1}%
\end{align}
while the projection of the field equations along $\Sigma_{d-2}$,
$\mathcal{E}_{m}=0$, reduces to%
\begin{align}
\epsilon_{m_{3}m_{4}\, \cdot\cdot\cdot \, m_{d}}  & \left[  \left(  \frac{l^{2}%
}{\mathcal{L}^{2}}-1-n(z^{2}-1)\right)  \bar{R}^{m_{4}m_{5}}\cdot\cdot
\cdot\bar{R}^{m_{d-1}m_{d}}\right.  \nonumber\\
& \left.  +A(\rho)\bar{R}^{m_{4}m_{5}}\cdot\cdot\cdot\bar{R}^{m_{d-3}m_{d-2}%
}\tilde{e}^{m_{d-1}}\tilde{e}^{m_{d}}\right]  =0\ ,\label{Em}%
\end{align}
with $A(\rho):=(1-n)\left(  z^{2}-1\right)  \left[  (z^{2}-3)\cosh^{2}%
(\rho)+z\sinh(2\rho)\tanh(z(\rho-\rho_{0}))\right]  $. It can be seen from eq.
(\ref{Em}) that at least one of the components of the $(d-3)$-form
\begin{equation}
\epsilon_{m_{3}m_{4}\cdot\cdot\cdot m_{d}}\bar{R}^{m_{4}m_{5}}\cdot\cdot
\cdot\bar{R}^{m_{d-1}m_{d}}\label{d-3-form}%
\end{equation}
must not vanish, otherwise the dynamical exponent $z$ would be arbitrary. This
is just a reflection of the fact that if (\ref{d-3-form}) vanished, actually
the metric would be undetermined since in this case the $g_{tt}$ component
becomes an arbitrary function. Therefore, the degeneracy in the metric is
removed requiring the dynamical exponent to be fixed as in eq.
(\ref{dynamical exponent}). Thus, eq. (\ref{Eo-E1}) is fulfilled by virtue of
eqs. (\ref{Em}) and (\ref{E0}).

\bigskip

In sum, for $z^{2}\neq1$, the metric of the base manifold $\Sigma_{d-2}$
fulfills the field equation%
\begin{equation}
\epsilon_{m_{3}m_{4}\cdot\cdot\cdot m_{d}}\bar{R}^{m_{4}m_{5}} \cdot\cdot
\cdot\bar{R}^{m_{d-3} m_{d-2}}\tilde{e}^{m_{d-1}}\tilde{e}^{m_{d}}=0\ ,
\label{Base manifold-m}%
\end{equation}
with an additional scalar condition%
\begin{equation}
\epsilon_{m_{3}m_{4}\cdot\cdot\cdot m_{d}} \bar{R}^{m_{3}m_{4}} \cdot
\cdot\cdot\bar{R}^{m_{d-2} m_{d-1}}\tilde{e}^{m_{d}}=0\ ,
\label{base manifold scalar}%
\end{equation}
provided at least one of the components of (\ref{d-3-form}) does not vanish.

\bigskip

The case $z^{2}=1$ allows much more freedom for the choice of base manifold
$\Sigma_{d-2}$, since its metric only fulfills the scalar condition
(\ref{base manifold scalar}), where at least one of the components of
(\ref{d-3-form}) is different from zero. This is in agreement with the results
found in \cite{DOT-Higher}. Note that for $z=1$, if the base manifold were
chosen as being locally isomorphic to the hyperbolic space of radius one, the
scalar condition (\ref{base manifold scalar}) would be trivially fulfilled,
but since (\ref{d-3-form}) also vanishes in this case the metric turns out to
be undetermined. Both conditions are satisfied for any compact and smooth base
manifold whose metric is given by the one of $\Sigma_{d-2}=S^{1}\times
H_{d-3}/\Gamma$, where the radius of the hyperbolic space $H_{d-3}$ is given
by $(2n-1)^{-1/2}$, and $\Gamma$ is a freely acting discrete subgroup of
$O(d-3,1)$.

\bigskip

In $d=5$ and $7$ dimensions the solution of the form (\ref{Lifshitz wormhole})
holds only for $z=1$. In five dimensions this is because the field equations
defined by eq. (\ref{Field Eqns}) correspond to the ones of the
Einstein-Gauss-Bonnet theory possessing a single maximally symmetric vacua of
squared radius given by $l^{2}$ --which cannot be degenerate unless
$\mathcal{L}^{2}=l^{2}$--, and hence $z=1$. Indeed, as explained in Refs.
\cite{DOT5,DOT6,DOT-Higher}, $\mathcal{L}^{2}=l^{2}$ is a necessary condition
for the existence of static asymptotically AdS wormholes in vacuum for the
Einstein-Gauss-Bonnet theory in $d\geq5$ dimensions. In seven dimensions the
reason is different. The field equations (\ref{Field Eqns}) correspond to a
cubic Lovelock theory. In this case, the field equation of the base manifold
metric given by (\ref{Base manifold-m}) means that $\Sigma_{5}$ has to be an
Euclidean Einstein manifold of negative scalar curvature, while the scalar
condition (\ref{base manifold scalar}) further restricts its geometry so that
$\Sigma_{5}$ has to be of constant curvature $-1$, i.e., $\bar{R}^{mn}=0$.
Therefore, the $(d-3)$-form in eq. (\ref{d-3-form}) vanishes, and hence the
$g_{tt}$ component of the metric becomes undetermined. As explained in the
previous paragraph, the conditions on $\Sigma_{5}$ are different for
$\mathcal{L}^{2}=l^{2}$ so that the asymptotically AdS solution
(\ref{Lifshitz wormhole}) with $z=1$ exists and it is well defined in seven dimensions.

Obstructions of the sort mentioned above do not apply for spacetimes of the
form (\ref{Lifshitz wormhole}) with $0\leq z^{2}\neq1$ in $d\geq9$ dimensions.
Nonetheless, finding an explicit metric for the base manifold that fulfills
the required conditions is not a straightforward task. This is left as an open problem.

\section{Asymptotically Lifshitz black hole}

\label{Black hole}

The theory described by (\ref{Field Eqns}) also admits a different solution,
whose metric is given by%
\begin{equation}
ds^{2}=l^{2}\left[  -\sinh^{2}(z\rho)dt^{2}+d\rho^{2}+\cosh^{2}(\rho
)d\Sigma_{d-2}^{2}\right]  \ ,\label{Black hole rho}%
\end{equation}
where $z$ is fixed in terms of the quotient of the curvature radii of the
maximally symmetric vacua as in eq. (\ref{dynamical exponent}). The solution
is well defined provided the bound (\ref{existence z}) is not saturated, i.e.,%
\begin{equation}
z^{2}>0\ \rightarrow\ \frac{l^{2}}{\mathcal{L}^{2}}>(1-n)\ .
\end{equation}
The base manifold metric $d\Sigma_{d-2}^{2}$ is independent of the coordinates
$t$, $\rho$ and fulfills the same conditions as the ones for the wormhole
described in the previous chapter; i.e., for $z^{2}\neq1$ the metric of
$\Sigma_{d-2}$ must solve the field equation (\ref{Base manifold-m}) with the
additional scalar condition (\ref{base manifold scalar}) provided at least one
of the components of (\ref{d-3-form}) does not vanish. The metric
(\ref{Black hole rho}) possesses an event horizon located at $\rho=0$, and it
describes an asymptotically Lifshitz black hole with dynamical exponent $z$,
since for $\rho\rightarrow+\infty$, the line element reduces to the one in eq.
(\ref{Lifshitz rho}). In terms of Schwarzschild-like coordinates,
$r=l\cosh(\rho)$ the metric reads%
\begin{equation}
ds^{2}=-4^{-z}\left(  \left(  \frac{r}{l}+\sqrt{\frac{r^{2}}{l^{2}}-1}\right)
^{z}-\left(  \frac{r}{l}+\sqrt{\frac{r^{2}}{l^{2}}-1}\right)  ^{-z}\right)
^{2}d\tau^{2}+\frac{dr^{2}}{\frac{r^{2}}{l^{2}}-1}+r^{2}d\Sigma_{d-2}%
^{2}\ ,\label{BH-r}%
\end{equation}
where the time coordinate has been rescaled as $\tau=2^{z-1}lt$ in order to
fit the standard form of Lifshitz spacetime (\ref{Pure Lifshitz}) in the
asymptotic region. The horizon is now located at $r=l$ and encloses the
singularity at the origin, $r=0$. Its Hawking temperature can be found
demanding regularity of the Euclidean solution at the horizon, and it is given
by%
\begin{equation}
T=\frac{z}{2^{z}\pi l}\ .\label{Temperature}%
\end{equation}
A number $m$ of additional singularities that shield the one at the origin,
are developed at $r=l\cos\left(  q\pi z^{-1}\right)  <l$, where $q\leq m$ is a
positive integer, provided the dynamical exponent fulfills $z>2m$. This can be
seen as follows. The inner region, $r<l$ is suitably covered by the patch
defined through $r=l\cos(\theta)$ with $0<\theta\leq\frac{\pi}{2}$, so that
the metric (\ref{BH-r}) reads%
\begin{equation}
ds^{2}=l^{2}\left[  -d\theta^{2}+\sin^{2}(z\theta)dt^{2}+\cos^{2}%
(\theta)d\Sigma_{d-2}^{2}\right]  \ .\label{BH-inner}%
\end{equation}
It is then apparent that the horizon ($\theta=0$) not only surrounds the
singularity at the origin ($\theta=\frac{\pi}{2}$), but also the additional
ones developed at $\theta=q\pi z^{-1}$. The singularity at the origin is
generically spacelike unless the dynamical exponent $z$ is an even integer so
that it becomes null\footnote{Curiously, requiring the singularity at the
origin to be null, quantizes the quotient $\frac{l^{2}}{\mathcal{L}^{2}}$ of
the curvature radii of the maximally symmetric vacua to be and even or odd
integer for odd and even $n$, respectively.}.

In the case of $z^{2}=1$, i.e., for $\mathcal{L}^{2}=l^{2}$, the solution is
asymptotically locally AdS and it can be extended to admit an integration
constant $r_{+}$ that parametrizes the horizon radius, so that the metric is
given by\footnote{As explained in \cite{SUSY-ground states},\ in eq.
(\ref{CS-BH}) the geometry of $\Sigma_{d-2}$ is arbitrary, but fixed by the
boundary conditions (see also \cite{DOT-Higher}), and it becomes further
restricted requiring (\ref{CS-BH}) to admit Killing spinors. This is an
extension of the solutions previously discussed in \cite{Dimensionally
continued, Cai:1998vy, ATZ}. Further aspects of this spacetime have been
discussed in Refs. \cite{Giribet:2006ec, Oliva:2010xn, Gonzalez:2010vv,
LopezOrtega:2010wx, Gonzalez:2011du, Mora:2004kb1, Mora:2006ka2,
Canfora:2007xs, Canfora:2010rh}.}%
\begin{equation}
ds^{2}=-\left(  r^{2}-r_{+}^{2}\right)  \frac{d\tau^{2}}{l^{2}}+\frac{l^{2}%
}{\left(  r^{2}-r_{+}^{2}\right)  }dr^{2}+r^{2}d\Sigma_{d-2}^{2}\ .
\label{CS-BH}%
\end{equation}
It would then be desirable exploring whether the asymptotically Lifshitz black
hole (\ref{Black hole rho}) could also be extended so as to admit an
integration constant.

\section{Further exact solutions}

\label{Further solutions}

The field equations (\ref{Field Eqns}) admit a wider class of asymptotically
Lifshitz solutions in vacuum, whose metric is given by%
\begin{equation}
ds^{2}=-\left(  a\left(  \frac{r}{l}+\sqrt{\frac{r^{2}}{l^{2}}+\gamma}\right)
^{z}+b\left(  \frac{r}{l}+\sqrt{\frac{r^{2}}{l^{2}}+\gamma}\right)
^{-z}\right)  ^{2}dt^{2}+\frac{dr^{2}}{\frac{r^{2}}{l^{2}}+\gamma}%
+r^{2}d\Sigma_{d-2}^{2}\ , \label{General solution}%
\end{equation}
where the dynamical exponent $z$ is again fixed as in eq.
(\ref{dynamical exponent}). Here $a$ and $b$ are integration constants, and
$\gamma$ can always be rescaled as $\gamma=\pm1,0$. The case of $\gamma=1$
generically leads to solutions with naked singularities, and so the remaining
cases of interest are discussed in what follows.

\bigskip

$\bullet$ $\gamma=-1:$

\bigskip

In this case, the black hole (\ref{BH-r}) is recovered from
(\ref{General solution}) with $a=-b=4^{-\frac{z}{2}}$. The wormhole metric
(\ref{Lifshitz wormhole}) can also be recovered from (\ref{General solution})
in the case of $a=\frac{1}{2}e^{-z\rho_{0}}$ and $b=\frac{1}{2}e^{z\rho_{0}}$,
followed by a change of coordinates given by $r\rightarrow l\cosh(\rho)$, and
$t\rightarrow lt$. This means that the region $r<l$ in (\ref{General solution}%
) can be consistently excised, so that two copies of the exterior region
$r>l$, with parameters $a=\frac{1}{2}e^{-z\rho_{0}}$, $b=\frac{1}{2}%
e^{z\rho_{0}}$ ($\rho>0$), and $a=\frac{1}{2}e^{z\rho_{0}}$, $b=\frac{1}%
{2}e^{-z\rho_{0}}$ ($\rho<0$) can be smoothly matched at $r=l$ ($\rho=0$) in
vacuum, without the need of introducing any kind of stress energy tensor at
the neck, as it would be necessary in the case of General Relativity. This is
a known feature of Lovelock gravity theories (see e.g., Refs. \cite{HTZ1,
HTZ2, Garraffo:2007fi, Garraffo:2010fn, Gravanis}).

\bigskip

A different solution is recovered from the metric (\ref{General solution})
with $b=0$, which after changing the coordinates as $r\rightarrow l\cosh
(\rho)$, and $t\rightarrow\frac{l}{a}t$, reads\footnote{The case $a=0$
corresponds to $b=0$ with $z\rightarrow-z$. \ Furthermore, it can be assumed
that $z>0$, since as for the previous wormhole solution, making $z\rightarrow
-z$ amounts to a total reflection in the radial coordinate.}%
\begin{equation}
ds^{2}=l^{2}\left[  -e^{2z\rho}dt^{2}+d\rho^{2}+\cosh^{2}(\rho)d\Sigma
_{d-2}^{2}\right]  \ , \label{asymwormhole}%
\end{equation}
where the base manifold $\Sigma_{d-2}$ fulfills the same conditions as the
solutions described above. This geometry describes a static wormhole with a
neck of radius $l$ located at $\rho=0$, and interpolates between
asymptotically Lifshitz spacetimes with different dynamical exponents given by
$z$ and $-z$, for $\rho\rightarrow\infty$, and $\rho\rightarrow-\infty$,
respectively. As it can be seen from (\ref{asymwormhole}), the gravitational
field pulls towards the asymptotic region $\rho\rightarrow-\infty$. In spite
of the fact that the curvature invariants are regular everywhere, it is simple
to verify that this spacetime is not geodesically complete, since null radial
geodesics can also reach $\rho=-\infty$ in a finite affine parameter, and the
warp factor of the base manifold blows up there.

\bigskip

$\bullet$ $\gamma=0:$

\bigskip

Making $b=0$ and $a=2^{-z}$ in (\ref{General solution}) the metric reads
\begin{equation}
ds^{2}=-\frac{r^{2z}}{l^{2z}}dt^{2}+\frac{l^{2}}{r^{2}}dr^{2}+r^{2}%
d\Sigma_{d-2}^{2}\ , \label{Pure Lifshitz gamma 0}%
\end{equation}
which corresponds to a Lifshitz spacetime with a nontrivial base manifold. In
the case of $z^{2}\neq1$, the geometry of $\Sigma_{d-2}$ fulfills the field
equation%
\begin{equation}
\epsilon_{m_{3}m_{4}\cdot\cdot\cdot m_{d}} \tilde{R}^{m_{4}m_{5}} \cdot
\cdot\tilde{R}^{m_{d-3} m_{d-2}}\tilde{e}^{m_{d-1}}\tilde{e}^{m_{d}}=0\ ,
\end{equation}
with an additional scalar condition, which reads%
\begin{equation}
\epsilon_{m_{3}m_{4}\cdot\cdot\cdot m_{d}} \tilde{R}^{m_{3}m_{4}}\cdot
\cdot\cdot\tilde{R}^{m_{d-2} m_{d-1}}\tilde{e}^{m_{d}}=0\ ,
\end{equation}
provided at least one of the components of the $(d-3)$-form given by%
\begin{equation}
\epsilon_{m_{3}m_{4}\cdot\cdot\cdot m_{d}} \tilde{R}^{m_{4}m_{5}}\cdot
\cdot\cdot\tilde{R}^{m_{d-1} m_{d}} \label{d-3-form-gamma=0}%
\end{equation}
does not vanish; else the $g_{tt}$ component of the metric becomes
undetermined, as it would the case if $\Sigma_{d-2}$ were chosen as a locally
flat spacetime.

For $z^{2}=1$ the base manifold metric has less restrictive conditions, since
it has to fulfill a scalar condition given by%
\begin{equation}
\epsilon_{m_{3}m_{4}\cdot\cdot\cdot m_{d}} \tilde{R}^{m_{3}m_{4}}\cdot
\cdot\cdot\tilde{R}^{m_{d-2} m_{d-1}}\tilde{e}^{m_{d}}=0\ ,
\end{equation}
where at least one of the components of (\ref{d-3-form-gamma=0}) does not vanish.

\bigskip

An additional curious solution is recovered once making $a=2^{-z}$, and
$b=b_{0}2^{z}$ in (\ref{General solution}). The metric is given by%
\begin{equation}
ds^{2}=-\left(  \frac{r^{z}}{l^{z}}+b_{0}\frac{l^{z}}{r^{z}}\right)
^{2}dt^{2}+l^{2}\frac{dr^{2}}{r^{2}}+r^{2}d\Sigma_{d-2}^{2}%
\ ,\label{Lifshitz-raro}%
\end{equation}
where the base manifold fulfills the same conditions as the pure Lifshitz
spacetime described above. This spacetime interpolates between an
asymptotically Lifshitz spacetime with dynamical exponent $z$, for
$r\rightarrow\infty$, and another Lifshitz spacetime of dynamical exponent
$-z$ at $r=0$, as it can be seen after a time rescaling given by $t\rightarrow
b_{0}^{-1}t$. For $z>0$, gravity pulls towards the surface defined by
$r=lb_{0}^{\frac{1}{2z}}$, and timelike geodesics turn out to be bounded,
since they neither reach the (repulsive) singularity at the origin nor the
asymptotic region.

\section{Summary and discussion}

\label{Summary and discussion}

A class of Lovelock theories selected by requiring that all but one of their
$n$ maximally symmetric vacua are AdS of radius $l$ and degenerate was
considered. The field equations (\ref{Field Eqns}) were shown to admit exact
static asymptotically Lifshitz wormholes and black holes in vacuum, given by
eqs. (\ref{Lifshitz wormhole}) and (\ref{Black hole rho}), respectively, for
$d=2n+1>7$ dimensions. The wormhole exists provided the curvature radius of
the nondegenerate vacuum is outside the range $\frac{l^{2}}{1-n}%
<\mathcal{L}^{2}\leq0$, and it connects two Lifshitz spacetimes with a
nontrivial geometry at the spacelike boundary. The dynamical exponent $z$ is
determined by the quotient of the curvature radii of the maximally symmetric
vacua as in eq. (\ref{dynamical exponent}), and then measures the deviation of
the non degenerate maximally symmetric vacuum of radius $\mathcal{L}$ with
respect to the $(n-1)$-degenerate AdS ones. The wormhole geometry is
geodesically complete and regular everywhere, and hence, as one approaches to
the inner region, the potentially divergent tidal forces appearing around the
origin of pure Lifshitz spacetime \cite{Horowitz-Ross, KLM, Copsey-Mann} are
circumvented by the presence of a throat.

In the case of $z=0$, the wormhole metric reduces to (\ref{wormhole z=0}) and
it connects two static universes at $\rho\rightarrow\pm\infty$. The
corresponding causal structures for $z>0$, and $z=0$ are depicted in Figs.
\ref{Fig}a and \ref{Fig}b, respectively.

It is worth pointing out that for General Relativity, as explained in
\cite{Morris:1988tu}, static wormhole solutions necessarily violate the
standard energy condition around the neck, while asymptotically Lifshitz
spacetimes also do for $z<1$ \cite{Lifshitz-energy conditions1,
Lifshitz-energy conditions2}. Remarkably, since the wormhole solution
(\ref{Lifshitz wormhole}) solves the field equations (\ref{Field Eqns}) in
vacuum for $z\geq0$, the spacetime is devoid of any kind of stress-energy
tensor everywhere, and hence no energy conditions can be violated. Therefore,
the results found for General Relativity do not apply for Lovelock gravity. It
would then be interesting to explore whether this spacetime, for a generic
dynamical exponent, could also be stable against scalar field perturbations,
as it is the case of its asymptotically AdS cousin for $z=1$
\cite{Correa:2008nq}.

It is known that wormholes raise some interesting puzzles in the context of
the gauge/gravity correspondence \cite{Witten:1999xp,Maldacena:2004rf,AHOP}.
Intriguing results along this line have been found in Refs.
\cite{Ali:2010qz,Arias:2010xg,Fujita:2011fn}.

As explained in Sec. \ref{Base maniflod geometry}, the geometry of the
spacelike boundary $\Sigma_{d-2}$ fulfills a curious set of conditions, which
for $z^{2}\neq1$, reduce to the field equation (\ref{Base manifold-m})
corresponding to the class of theories discussed in \cite{BH-Scan}, together
with the additional scalar condition (\ref{base manifold scalar}). An
additional condition, which requires that at least one of the components of
(\ref{d-3-form}) does not vanish, has to be imposed; otherwise, the $g_{tt}$
component of the metric would be an arbitrary function, as it is the case of
asymptotically Lifshitz solutions for Lovelock theories in vacuum previously
found in the literature. Finding an explicit Euclidean metric that fulfills
the required conditions turned out to be not a simple task and it is then left
as an open problem. Indeed, product manifolds of constant curvature solving
both (\ref{Base manifold-m}) and (\ref{base manifold scalar}) for certain
fixed radii, simultaneously make the $d-3$-form (\ref{d-3-form}) to vanish,
making the $g_{tt}$ component of the metric to be undetermined. Different
classes of solutions with nontrivial geometries at the boundary for Lovelock
theories in vacuum have also been considered in e.g. Refs. \cite{Cai:1998vy,
ATZ, Cai, DOT5, DOT6, DOT-Higher, Maeda:2011ii,Camanho:2011rj,BCGZ}.

\bigskip

The asymptotically Lifshitz black hole solution whose metric is given by
(\ref{BH-r}) possesses the same dynamical exponent (\ref{dynamical exponent}),
with a fixed event horizon that surrounds the singularities, and a Hawking
temperature given by (\ref{Temperature}). It would also be interesting to
explore the stability of this solution, its entropy, as well as whether it
could be extended so as to admit an integration constant parametrizing the
horizon radius.

\bigskip

The wormhole (\ref{Lifshitz wormhole}) and the black hole
(\ref{Black hole rho}) were found to belong to a wider class of asymptotically
Lifshitz solutions in vacuum, given by the metric (\ref{General solution}).
This class was shown to include pure Lifshitz spacetimes with a nontrivial
geometry at the spacelike boundary, given by (\ref{Pure Lifshitz gamma 0}),
wormholes that interpolate between different asymptotically Lifshitz
spacetimes with dynamical exponents given by $z$ and $-z$, as in
(\ref{asymwormhole}), and the solution (\ref{Lifshitz-raro}) that interpolates
between an asymptotically Lifshitz spacetime with dynamical exponent $z$, and
another Lifshitz spacetime of dynamical exponent $-z$ at the origin.

\bigskip

As a final remark, it would be worth exploring whether the class of Lovelock
theories considered here could be widened so as to admit well-behaved
asymptotically Lifshitz solutions in vacuum.

\bigskip

\textit{Acknowledgments.} We thank Hern\'{a}n Gonz\'{a}lez, Mokhtar
Hassa\"{\i}ne, Hideki Maeda, Alfredo P\'{e}rez, and specially to Cristi\'{a}n
Mart\'{\i}nez and David Tempo for useful and enlightening discussions. This
work has been partially funded by the Fondecyt grants N$%
%TCIMACRO{\U{b0}}%
%BeginExpansion
{{}^\circ}%
%EndExpansion
$ 1085322, 1095098, and by the Conicyt grant ACT-91: \textquotedblleft
Southern Theoretical Physics Laboratory\textquotedblright\ (STPLab). J.M.
thanks Conicyt for financial support. The Centro de Estudios Cient\'{\i}ficos
(CECs) is funded by the Chilean Government through the Centers of Excellence
Base Financing Program of Conicyt.

\appendix

\section{Absence of the $k$-th power of the curvature in the action and a
special class of dynamical exponents}

\label{Appendix}

Let us consider the solutions in vacuum discussed above in the case for which
the dynamical exponent lies within the range $0\leq z^{2}\leq1-1/n$. Here we
show that there is a special class, defined by%
\begin{equation}
z^{2}=1-\frac{1}{n-k}\ ,
\end{equation}
where $k<n$ is a non negative integer, that corresponds to Lovelock theories
of the form (\ref{Field Eqns}) being such that the $k$-th power of the
curvature is absent in the field equations, and hence in the action. By virtue
of (\ref{dynamical exponent}), the field equations (\ref{Field Eqns}) read%
\begin{equation}
\epsilon_{aa_{2}a_{3}\cdot\cdot\cdot a_{d}}\bar{R}^{a_{2}%
a_{3}}\cdot\cdot\cdot\bar{R}^{a_{d-3}a_{d-2}}\left(  R^{a_{d-1}a_{d}}%
+\frac{n(z^{2}-1)+1}{l^{2}}e^{a_{d-1}}e^{a_{d}}\right)  =0\ ,
\end{equation}
which is equivalent to%
\begin{equation}
\epsilon_{aa_{2}a_{3}\cdot\cdot\cdot a_{d}} \sum
_{k=0}^{n}\frac{n!}{k!(n-k)!} l^{2(k-n)}\left[  (n-k)(z^2-1)+1\right]
\bar{R}^{a_{2}a_{3}}\cdot\cdot\cdot\bar{R}^{a_{2k}a_{2k+1}}e^{a_{2k+2}}%
\cdot\cdot\cdot e^{a_{d}} =0\ .
\end{equation}
Therefore, it is apparent that the $k$-th power of the curvature is absent
provided%
\[
z^{2}=1-\frac{1}{n-k}\ .
\]
In particular, for $z^{2}=\frac{n-2}{n-1}$, the Einstein-Hilbert term is
absent in the action.

\end{document}